\documentclass[runningheads]{llncs}
\usepackage[T1]{fontenc}
\usepackage{graphicx}
\usepackage[colorlinks=true, linkcolor=blue, citecolor=blue, urlcolor=blue]{hyperref}
\usepackage{ulem}
\usepackage{enumitem}
\usepackage{tabularx}
\usepackage{booktabs}
\usepackage{soul}

\begin{document}
\title{A Cooperative Approach for Knowledge-based Business Process Design in a Public Authority\thanks{This work has been partially supported by the PNRR MUR project PE0000013-FAIR CUP B53C22003630006, Next Generation EU program, Italian Government.}}
\titlerunning{A Cooperative Approach for Knowledge-based BP Design}
%
\author{Mohammad Azarijafari\inst{1}\orcidID{0000-0002-3018-7000} \and
Luisa Mich\inst{1}\orcidID{0000-0002-0018-6883} \and
Michele Missikoff\inst{2,*}\orcidID{0000-0002-7972-5201} \and
Oleg Missikoff\inst{3}}
\authorrunning{M. Azarijafari et al.}
%
\institute{University of Trento, Department of Industrial Engineering, Via Sommarive, 9, 38123, Trento, Italy\\
\email{\{mohammad.azarijafari,luisa.mich\}@unitn.it} \and
National Research Council (CNR), Istituto di Analisi dei Sistemi ed Informatica (IASI) “Antonio Ruberti”, Via dei Taurini, 19, 00185, Rome, Italy\\
\email{\textsuperscript{*}michele.missikoff@iasi.cnr.it} \and
Sapienza University of Rome, Department of European, American And Intercultural Studies, Piazzale Aldo Moro, 5, 00185, Rome, Italy\\
\email{oleg.missikoff@uniroma1.it}}
\maketitle              
\begin{abstract}
Enterprises are currently undergoing profound transformations due to the unpostponable digital transformation. Then, to remain competitive, enterprises must adapt digital solutions, transforming their organisational structures and operations. This organisational shift is also important for small and medium-sized enterprises. A key innovation frontier is the adoption of process-oriented production models. This paper presents a knowledge-based method to support business experts in designing business processes. The method requires no prior expertise in Knowledge Engineering and guides designers through a structured sequence of steps to produce a diagrammatic workflow of the target process. The construction of the knowledge base starts from simple, text-based, knowledge artefacts and then progresses towards more structured, formal representations. The approach has been conceived to allow a shared approach for all stakeholders and actors who participate in the BP design.

\end{abstract}

\section{Introduction}\label{sec1-introduction}

In today’s increasingly interconnected and competitive global economy, the ability of enterprises to collaborate effectively has become a critical success factor. This is particularly true for small and medium-sized enterprises (SMEs), which often face significant limitations in terms of resources, production capacity, and market reach \cite{audretsch2023collaboration}. While individually these firms may struggle to meet the complex demands of modern production systems, their collective potential can be greatly enhanced through the coordination of different functions and actors within the organisation.

Cooperative work environments allow these functions to pool resources, share expertise, and optimize operations within a structured framework. For SMEs, this approach provides a strategic path to scale up activities, engage in more efficient workflows, and increase resilience in the face of market fluctuations. By promoting collaboration, they can address production challenges that would otherwise be unattainable, fostering innovation, improving efficiency, and enhancing overall competitiveness \cite{xiangfeng2025determinants}.

To achieve such cooperation, an important prerequisite is a common understanding and sharing of knowledge about the application context and the business operations. To this end, it is essential to carry out an analysis of the business processes and create formal representations beyond simple textual descriptions \cite{de2023formal}. Such a formal representation is the first step in designing a knowledge base to support structured and cooperative business process design.

In this paper, we introduce the Business Process Analysis (BPA) Canvas, a methodological framework aimed at guiding the design of business processes in a cooperative environment. The proposed framework represents the first phase of a more encompassing digital transformation effort. The design of a business process typically starts with the analysis of the state of practice, referred to as As-Is analysis, and then evolves with the identification of needed improvement towards the new solutions that will drive the construction of the new business process, known as the To-Be configuration.

To demonstrate the practical application of the proposed methodology, we apply the BPA Canvas method to a real-world case study from the Malawi Energy Regulatory Authority (MERA). This public authority is responsible for issuing licences to enterprises that act as intermediaries in selling energy to end customers. This case provides a concrete scenario to validate the methodology and show how knowledge can be collected, structured and shared within a cooperative business environment.

The remainder of the paper is structured as follows. Section \ref{sec2-related-work} presents the related work, while Section \ref{sec3-BPA-convas} illustrates the structure of BPA Canvas methodology, outlining its components and knowledge segments. Section \ref{sec4-mera-canvas} describes a first application of the method to the MERA licensing process and Section \ref{sec5-bp-diagram} reports a graphical representation of the BP. Finally, Section \ref{sec6-conclusion} concludes the paper and outlines directions for future research.

\section{Related Work}\label{sec2-related-work}

A variety of approaches have been proposed to support Business Process Analysis (BPA) aiming at digital innovations and process automation. Goni et al. examined sustainable business models and proposed a framework for embedding sustainability within business processes \cite{goni2021sustainable}. Ahmad and Van Looy identified key trends in the digital transformation of Business Process Management (BPM), including strategic alignment and business-IT integration \cite{ahmad2020business}. Schüler and Alpers categorised methods for automated process model generation and outlined challenges related to data quality and model applicability \cite{schuler2024state}. Similarly, Weinzierl et al. reviewed machine learning applications across the BPM lifecycle, highlighting their potential in optimisation and decision support \cite{weinzierl2024machine}.

In the area of Business Process Knowledge Bases (BPKBs), several frameworks have been developed to formalise and structure process knowledge. Motta et al. proposed a web-based knowledge management system that combines performance metrics, process structuring, and benchmarking \cite{motta2007performing}. Missikoff introduced a methodology for BPA based on Model-Driven Architecture, transforming textual process descriptions into ontologies \cite{missikoff2021knowledge}. Ligeza and Potempa introduced an Artificial Intelligence (AI) framework that combines Business Process Model and Notation (BPMN) \cite{chinosi2012bpmn} with Business Rules to enhance semantic accuracy \cite{ligkeza2012artificial}. Azarijafari et al. introduced an Agentic AI approach to BP Development, shifting from traditional task-oriented workflows to a goal-driven model \cite{azarijafari2025agentic}.

While previous studies have contributed valuable insights into business process modelling, they often focus on isolated aspects or specific technologies without offering an integrated methodological framework. Furthermore, the existing proposals often resulted to be cumbersome and not sufficiently user-friendly to gain real world acceptance. In contrast, with the knowledge-based approach of BPA Canvas we conceived a method to be easily adopted by business people without a technical background. Furthermore, BPA Canvas is characterised for a collaborative analysis and design of business processes, to support the development of workflows in complex industrial ecosystems.

\section{The BPA Canvas Methodology}\label{sec3-BPA-convas}

In this section, we present the BPA Canvas methodology that in its full representation is organised in eight knowledge segments. The first objective of the BPA Canvas is to facilitate the collection of the knowledge about a given BP in a collaborative, shared fashion. In this way, we ensure a large involvement of all the stakeholders and players that collaboratively participate in the specification of the BP. Such a wide participation contributes to improving the quality of the analysis report, reducing missing information, ambiguities and imprecision. The BPA Canvas also represents the structure of the shared BP knowledge base (BPKB), a repository organised accordingly. A collection of BPKBs are able to model a decentralised production ecosystem and the set of interconnected BPs taking place therein.

The eight knowledge segments of BPA Canvas hold different kinds of knowledge artefacts, i.e., knowledge models built observing the given business domain. Such artefacts can assume various forms, with different levels of detail and formality. In particular, they can be represented as: 

\begin{enumerate}[label=\roman*.]

    \item \textbf{Plain text.} A narrative form of knowledge representation.
    \item \textbf{Structured text.} Like itemised lists using bullet points to collect and organise short statements.
    \item \textbf{Tables.} Typically, systematic visualisation of knowledge artefacts in a tabular format.
    \item \textbf{Diagrams.} Graphical representations of knowledge artefact using a defined standard.
    \item \textbf{Formal representation.} Modelling the business domain by means of a BP Ontology

\end{enumerate}

The eight knowledge segments of the BP canvas are:

\begin{enumerate}

    \item \textbf{BP Signature.} The first knowledge artefact, in the form of a simple table, aimed at providing a synthetic description of the business process.
    \item \textbf{BP Statement.} A preliminary plain text description of the business process, and its business scenario, described at the intensional level and in general terms.
    \item \textbf{BP User story.} One or more plain text descriptions of exemplar executions of the BP at an extensional level. In essence, it represents one or more instance of the BP Statement.

\end{enumerate}

The first three segments represent the ‘BPA Canvas Lite’ that is recommended for simple processes and, in particular, for SMEs. The following segments are reported for sake of completeness, but they will not be used in this paper.

\begin{enumerate}
    \setcounter{enumi}{3}
    
    \item \textbf{Actor, Process, Outcome (APO) Matrix.} It represents a first operational excerpt of the BP Statement. Please note that here we use the term ‘process’ in a general sense, referring to any type of action, operation, activity, or function.
    \item \textbf{BP Glossary.} A collection of terms, with their descriptions, that characterise the BP domain.
    \item \textbf{OPAAL Lexicon.} This is a structured terminology that provides a first semantic tagging of the key terms used in the previous segments, based on categories of Object, Process, Actor, Attribute, Link.
    \item \textbf{UML Class Diagram.} A set of diagrams representing the structural relationships of the entities and the BP knowledge collected so far.
    \item \textbf{BP Ontology.} A formal representation of the analysed business process and its context. It is the final knowledge artefact of the methodology.

\end{enumerate}

The collected knowledge is finally used to draw the BP workflow, i.e., a diagram representation that includes actors, tasks, and their sequencing. To this end we adopt the BPMN graphical standard.

\section{Application of the BPA Canvas in MERA}\label{sec4-mera-canvas}

MERA is the institution responsible for issuing the necessary licences to enterprises that want a connection to the electricity grid in Malawi. The licensing process may concern also a group of SMEs that wants to coordinate to obtain a collective licence. The primary objective of this process is to ensure that licences are granted solely to qualified and compliant individuals or enterprises. This approach safeguards the integrity, compliance, and safety standards within Malawi’s energy sector. The meticulous and systematic design of the licensing process ensures that all regulatory requirements are fulfilled prior to licence issuance. In essence, the process starts when an applicant submits to MERA the form requesting the access to the electricity grid. The application form is verified to check the validity of the provided data and then an additional level of verification consists in the interview of the applicant. Finally, when all the data are complete and correct, the applicant is invoiced and, upon payment, the licence is issued. 

To demonstrate the practical use of the BPA Canvas, we describe its application to the MERA Licensing BP. As anticipated, we adopt a simplified version, BPA Canvas Lite, focusing on the first three segments of the BPA Canvas, which are \textbf{BP Signature}, \textbf{BP Statement}, and \textbf{BP User Story}.

\subsection{BP Signature}\label{subsec4-1}

The BP Signature represents the initial knowledge artefact in the BPA Canvas methodology. It provides a high-level overview of the business process by capturing its essential components in a concise, tabular format. Table \ref{tab:mera} provides the BP signature corresponding to the MERA licensing process, detailing the principal actors, involved objects, input elements, outputs, and the main objective of the business process.


\begin{table}[h!]
\caption{BP Signature for the MERA Licensing Process}\label{tab:mera}
    \begin{tabularx}{\textwidth}{@{}l |X@{}}
        \toprule
        \textbf{BP Name}     & MERA Licence Application \\ \midrule
        \textbf{Key Actors}  & Applicant, MERA Team, Licensing Office, CEO \\ \midrule
        \textbf{Key Objects} & Application Form, Application Form Integration Request, Application Form Integrated, Invoice, Licence Payment Receipt, Licence Signed, Licence Registration Number, Application Submission Check List, Site Inspection Check List \\ \midrule
        \textbf{Input}       & Application Form Submitted \\ \midrule
        \textbf{Objective}   & To process an issue energy-related licences efficiently and in compliance with regulatory standards. \\ \midrule
        \textbf{Output}      & Issued Licence \\
        
        \bottomrule

    \end{tabularx}
\end{table}


In BPA Canvas all the labels must be defined in the BP Glossary (omitted in this Lite version), a list of terms with definitions. This knowledge segment is particularly important to unify the terminology and avoid possible misunderstanding, especially in business ecosystems with multiple cooperating enterprises.

\subsection{BP Statement}\label{subsec4-2}

The BP Statement is a natural language description of the key phases of the BP, identifying the main actors, activities and objects, typically documents and database entries. It represents an important part of the BPKB. As it is expressed in natural language, it is easy to be constructed in a collaborative way by business experts in the enterprise domain, without the help of technical people.

The BP Statement for the MERA Licence BP is as follows:

\begin{enumerate}
    \item \textbf{Submission and Registration of Application}
    \begin{itemize}
        \item The licensing process commences when an applicant submits a fully filled-out application form along with the required supporting documents to the Licensing Office.
        \item Upon receipt, the Licensing Office registers the application and forwards it to the MERA Team for administrative check.
    \end{itemize}
    
    \item \textbf{Verification and Integration of Application Data}
    \begin{itemize}
        \item The MERA Team conducts a thorough review to confirm the completeness of the submitted data. 
        \item If any information is missing, the applicant is promptly notified and requested to submit the required details. 
        \item Once the application is verified as complete, it advances to the subsequent step.
    \end{itemize}
    
    \item \textbf{Board Committee Interview and Evaluation}
    \begin{itemize}
        \item The Board Committee conducts an interview with the applicant to evaluate its qualifications and capabilities relevant to the requested licence. 
        \item The outcome of this interview is carefully assessed to ensure compliance with necessary criteria.
    \end{itemize}

    \item \textbf{Fee Invoicing and Payment Confirmation}
    \begin{itemize}
        \item Applicants who successfully pass the evaluation are issued an invoice for the applicable licensing fee. 
        \item The process continues after the Licensing Office verifies receipt of payment.
    \end{itemize}

    \item \textbf{CEO Approval and Licence Issuance}
    \begin{itemize}
        \item Following the payment verification, the application is forwarded to the CEO for final approval and signature. 
        \item Upon receiving the CEO’s approval, the Licensing Office registers the licence, allocates a registration number, and officially issues the licence to the applicant.
    \end{itemize}
    
\end{enumerate}

In summary, Malawi's energy sector licensing process includes rigorous checks and evaluations aimed at granting licences exclusively to applicants who satisfy essential qualifications and regulatory criteria, thereby ensuring sector safety and reliability.

\subsection{BP User story}\label{subsec4-3}

In this segment of the BPA Canvas, we report a real-world example of the Malawi Licence BP, which is compliant with the BP Statement. This is also a way to double-check if the description reported in the BP Statement is correct. For this purpose, we describe the case of Mr. Kwame Banda, who applies for a business licence under the Malawi licensing process.

\begin{enumerate}
    \item \textbf{Submission and Registration of Application}

    \noindent
    Mr. Kwame Banda seeks to obtain a licence to operate as a MERA-certified electric installer in Malawi. He downloads the Licensing Application Form from the MERA website, carefully fills it out, and gathers all necessary supporting documents, including his business plan and proof of technical qualifications.
    
    After ensuring that the application is complete, Mr. Banda submits the form and documents to the Licensing Office. The Licensing Office receives his application, logs it into its system, and assigns it a reference number. The application is then forwarded to the MERA Team for further review.\\
    
    \item \textbf{Verification and Integration of Application Data}

    \noindent
    The MERA Team begins their review of Mr. Banda's submitted application. They check to ensure that all required fields have been filled in and that all necessary documents are attached. During their review, the MERA Team identifies that Mr. Banda’s is incomplete and needs to be completed.
    
    The team promptly informs Mr. Banda of the missing information and requests that he provides the missing competencies. Mr. Banda quickly gathers the required data and submits them to MERA, which then updates the application. Being all data now complete, the application moves forward to the next stage.\\
    
    \item \textbf{Board Committee Interview and Evaluation}

    \noindent
    Mr. Banda is scheduled for an interview with the MERA Board Committee. During the interview, the committee assesses whether his qualifications and experience are adequate to operate as a MERA-certified electric installer. They ask detailed questions about his technical expertise and his ability to comply with regulatory standards.
    
    Following the interview, the Board Committee deliberates and concludes that Mr. Banda possesses the necessary qualifications and competence to be issued a licence. The committee’s favourable evaluation is then recorded in the system, and the process proceeds to the next step.\\

    \item \textbf{Fee Invoicing and Payment Confirmation}

    \noindent
    With the Board Committee's approval, Mr. Banda receives an invoice from the Licensing Office, detailing the licence fee he must pay. Mr. Banda promptly makes the payment through an electronic transfer.
    
    The Licensing Office verifies that the payment has been received and correctly processed. Once the payment is confirmed, the application is sent to the CEO for final approval.\\
    
    \item \textbf{CEO Approval and Licence Issuance}

    \noindent
    The CEO of MERA reviews the application, the Board Committee's evaluation, and the payment confirmation. Satisfied with the thoroughness of the process, the CEO signs the licence, granting Mr. Banda the official permission to operate as a MERA licencee.

    The Licensing Office then registers the licence, assigns it a unique registration number, and issues the official licence document to Mr. Banda. Mr. Banda is notified that his licence has been approved and issued, and he receives the document, complete with the registration number.

\end{enumerate}

Ultimately, through this detailed process, Mr. Kwame Banda successfully navigates the MERA licensing procedure, ensuring that all regulatory requirements are met. The careful scrutiny and systematic steps ensure that only qualified individuals like Mr. Banda are granted licences, thus maintaining the integrity and safety of Malawi's energy sector.

\section{The BP Diagram}\label{sec5-bp-diagram}

\begin{figure}[h]
\centering
\includegraphics[width=\textwidth]{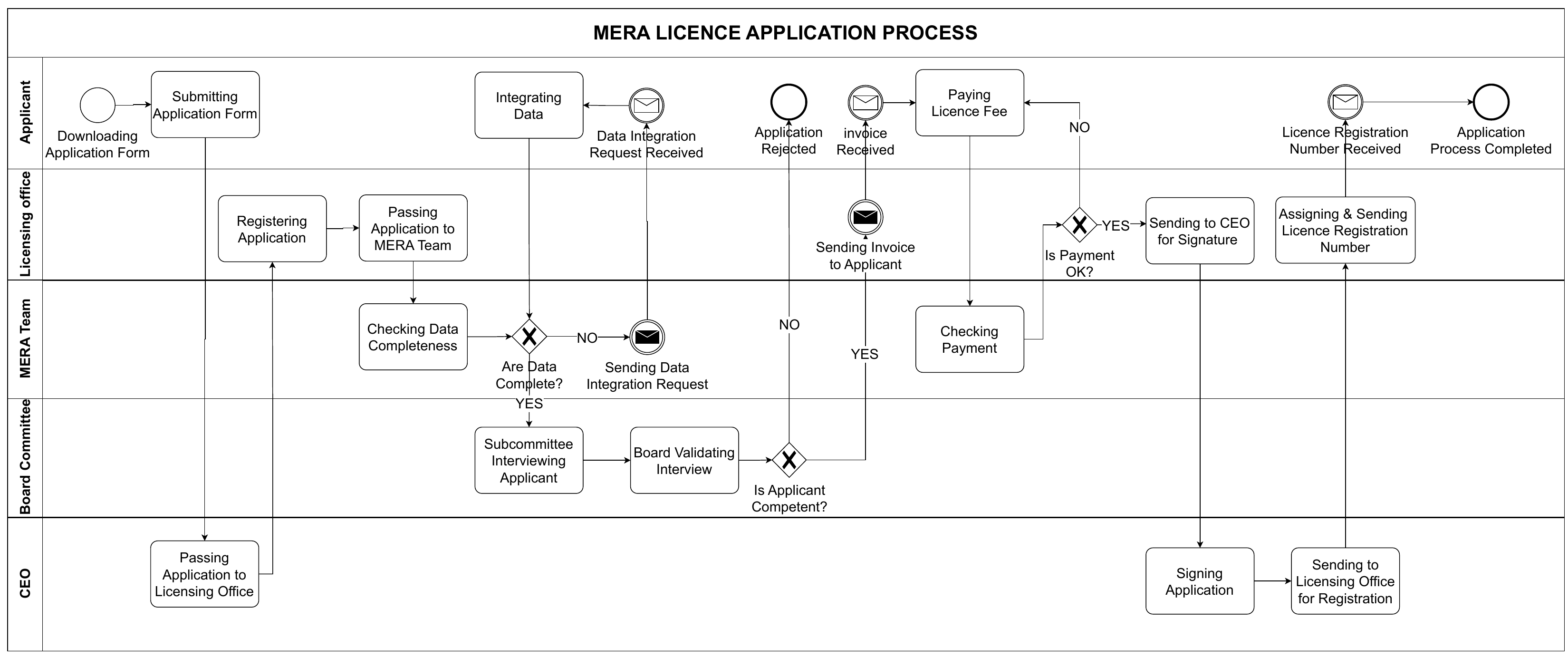}
\caption{BPMN representation of the MERA licencing process}\label{fig:mera}
\end{figure}

The BP diagram is a graphical representation of the process described in the BP Statement. As illustrated in Figure \ref{fig:mera}, the BP is represented using a BPMN diagram as a pool in which actors are represented as lanes, and tasks are connected by directed arcs to indicate the flow of activities. Different nodes can represent tasks (boxes with rounded corners), decision points (diamonds), or events (circles). For a more exhaustive introduction, please refer to the wide literature and tutorials.


In the diagram we see five lanes, corresponding to the key actors of the BP, namely: Applicant, Licensing Office, MERA Team, Board Committee, and CEO. The BP starts with the Applicant filling out the request form and ends with the rejection, in case of incompetent Applicant, or a positive completion.

The BPMN diagram visually represents the key stages of the MERA licensing process as described in the previous BPA Canvas segments. It serves both as a rigorous representation and a practical tool for communicating the structure and the logic of the workflow to various stakeholders. By aligning the diagram with the BP Signature, Statement, and User Story, the model ensures coherence between textual and visual artefacts, thereby enhancing understanding, and the potential for future automation. This graphical abstraction also provides a foundation for the representation of multiple processes belonging to a number of partner SMEs that are part of a broader decentralised production ecosystem.

\section{Conclusion}\label{sec6-conclusion}

In today’s increasingly interconnected and competitive global economy, the ability of enterprises to collaborate effectively has become a critical factor for success. This is particularly true for SMEs, which often face limitations in resources, market access, and technological capabilities. By fostering extended cooperation, SMEs can collectively address production challenges that would be unattainable individually. This collaboration fosters innovation, enhances efficiency, and strengthens their overall competitiveness. However, to make such cooperation effective, it is essential to first establish a shared understanding of the business context in which these enterprises operate, and BPs therein.

This study introduced a knowledge-based methodology, namely, the BPA Canvas, as a structured and cooperative framework for business process analysis and representation. By enabling a common representation of business processes and involving key stakeholders in the knowledge modelling phase, the methodology contributes to produce high quality process design, reduced ambiguity, and more informed decision-making. Through the case study of the MERA licensing process, we demonstrated how the BPA Canvas, which represents the foundation for cooperative process design, can be practically applied to capture operational knowledge and align goals.

Looking ahead, future work will explore how the outputs of the BPA Canvas can be operationalized using AI-based agents to support autonomous process execution within cooperative production environments. In particular, we aim to investigate how agentic AI systems can interpret, reason about, and act upon the business knowledge embedded in the BPA Canvas. Moreover, future extensions of this work may explore how the proposed methodology can be adapted to support decentralised process design and cross-organisational collaboration, especially in scenarios involving distributed actors and shared operational goals. This would represent a significant advancement toward building self-adaptive and intelligent production ecosystems that can respond dynamically to evolving business needs, further empowering SMEs in their digital transformation journey.



%
%
%
\bibliographystyle{splncs04}
\bibliography{mybibliography}
%




\end{document}